# The Integrated Approach to ERP: Embracing the Web

Sergey V. Zykov
ITERA International Group of Companies
Moscow, Russia
szykov@itera.ru

**Abstract**

Integrated approach to enterprise resource planning (ERP) software design and implementation can significantly improve the entire corporate information infrastructure and it helps to benefit from power of Internet services. The approach proposed provides for corporate Web portal integrity, consistency, urgency and front-end data processing. Human resources (HR) ERP component implementation is discussed as an essential instance.

## 1. Introduction

Frequent priority changes in conglomerate development require fast and flexible adaptability of management to match rapidly changing market conditions. Such adaptability should be based on strategic software integration and its connection to Internet, especially for comprehensive ERP.

During the two recent decades, the data models (DM) and architectures underlying software development process have been changed significantly to support object methodologies and interoperability. Since early 90's, rather primitive file-server systems supporting relational [3] DM have evolved into comprehensive ERP systems based on extendable relational DM supporting object-oriented DBMS. Attempts of enterprise application integration have also been undertaken [2,4-6].

The main objectives of the paper are development of integrated data and metadata model, application of the model for extending corporate Internet presence, formal approach to Web-enhanced ERP construction and overview of improved ERP implementation process. Research methods meeting the problem domain specific features are based on a creative synthesis of fundamental statements of lambda calculus [11], categories [1] and semantic networks [9].

The data model introduced provides an open, integrated, problem-oriented, event-driven data and metadata management of dynamic, heterogeneous, weak-structured problem domains in a more adequate way than previously known ones. The model suggested allows to generate system architecture solutions for open, distributed, interoperable environments supporting front-end, multi-purpose data warehousing and Internet-publishing on the basis of CORBA, UML and business-process reengineering (BPR) technologies.

## 2. Related Works

Papers [1,3,9-12] provide rigorous mathematics foundation and solid theoretical background for notations used. Relational DBMS and weak-structured document solutions are cross-examined. Papers [13,14] provide a mathematically rigorous overview of object-oriented systems development and describe a number of promising implementations. ERP generation history is discussed in [8,15].

## 3. Architecture and Interface Requirements

Specific features of the problem domain require support of dynamic multi-level personnel restructuring process with multi-alternative assignment-based comprehensive estimation of enterprise activity. Interface requirements set should allow dynamic variation of mandatory input fields, flexible access rights differentiation and non-interruptible data integrity support. In respect of architecture, the system should provide interoperability, expandability, flexible adjustment to problem domain changes and easy data/metadata correction (including rollback mechanisms).

## 4. The Integrated Data and Metadata Model

### 4.1 The Data Object Model

Mathematical formalisms existing for problem domains are not fully adequate to dynamics and statics semantics. Moreover, current methods of CASE-and-RAD enterprise applications development do not result in solutions of a wide application range; commercial ERP implementations do not provide sufficient flexibility of heterogeneous data handling. According to research results of personnel management problem domain, a computational DM based on object calculus has been built. The model is an innovative synthesis of finite sequence, category and semantic network theories.

Data objects (DO) of the DM can be represented as follows: DO = < concept, individual, state >, where a *concept* is understood as a collection of functions with the same definition area and the same value range. An *individual* implies an essence selected by a problem domain expert, who indicates the identifying properties. *State* changes simulate dynamics of problem domain individuals.

Compared to research results known as yet, the DM suggested features more adequate dynamics mapping for heterogeneous problem domains. The DM also benefits better support for problem-oriented integrated data management. In architecture and interface aspects the DM provides straightforward iterative design of open, distributed, interoperable HRIS based on UML and BPR methodologies. As far as implementation part is concerned, multi-repository information processing of heterogeneous problem domains is supported. Thus, front-end data access is provided which is based on event-driven procedures and dynamic SQL technologies.

The computational model suggested is based on the two-level conceptualization scheme [13]. *Conceptualization* implies a process of establishing relationship between problem domain concepts.

Individuals h, according to the assigned types T, are united in assignment-dependent collections, thus forming variables of sort $H_T(I) = \{h \mid h : I \rightarrow T\}$. This formalism is used to simulate problem domain dynamics.

When fixing data model individuals, uniqueness of individualization of data object d from problem domain D by means of the formula $\Phi$ is required:

$\| Ix \Phi (x) \| i = d \Leftrightarrow \{d\} = \{d \in D \mid \|\Phi(d)\| i = 1\}$.

### 4.2 The Metadata Object Model

Let us introduce a compression principle for the computational data object model

$C = Iy: [D] x : D(y(x) \leftrightarrow \Phi ) = \{x : D \mid \Phi\}$

that allows to apply the model to concepts, individuals and states separately, as well as to data objects as a whole. The suggested computational metadata model expands traditional ER-model [3] by adding the following compression principle:

$x^{j+1} Iz^{j+1}: [\ldots[D]\ldots] \forall x^j: [\ldots[D]\ldots] (z^{j+1}(x^j) \Phi^j)$,

where $z^{j+1}$, $x^{j+1}$ – metadata predicate characters in relation to level j, $x^j$ – individual and $\Phi^j$ – DO definition language construction of level j.

The integrated model for objects of data, metadata and states is characterized by scalability, expandability, metadata encapsulation and transparent visualization. Expandability, adequacy, neutrality and semantic correctness of the formalism introduced provide problem-oriented software design with adequacy maintenance at every stage of the implementation process.

Semantics of computational model of objects of data, metadata and states can be adequately and uniformly formalized by means of typed λ-calculus, combinatory logic, and semantic network-based scenario description.

### 4.3 Model Application for Web Site Enhancement

#### Enhancing Integrated ERP with Data View Representation Model

Let us consider the following parameters of Internet user appearance and behavior: data access rights, personal preferences (fonts, color settings, etc.), Web browser settings (links, cache, history, etc.) and data access device (Web TV, PDA, mobile phone, terminal, etc.) profile.

Let us assume that A and B are sets. Let $B^A$ stand for the mapping from A to B: $B^A = \{f \mid f : A \rightarrow B\}$.

Let us match $B^A$ with the $\|\circ\|$ evaluation function:

$\|\circ\| = \{f \mid f B^A \times A \rightarrow B\}$.

Thus, $\|\circ\| = (<f, x>) = f(x)$, so $\|<f, x>\| = f(x)$.

Now let us build the semantics network language model. Let us consider an ordered pair of DO of the form $L=<R,C>$, where $R=\{R_1,R_2, \ldots\}$ is a dyadic predicate symbols set and $C=\{C_1,C_2, \ldots\}$ is a set of constants. Therewith, the atomic formulae of the model suggested correspond to simple frames, and terms denote problem domain individuals. Let us construct a frame evaluation procedure using the introduced evaluation function $||\circ||$.

Now let us consider an example of user profile evaluation procedure based on the suggested data model. Let the F functional denote the most general class of users. Let the assignment s={high resolution graphics, multimedia} account for user customary settings. Let F(s) stand for the set of users, for whom the customary settings are restricted to high resolution graphics and multimedia.

Let the assignment p={registered, unregistered, corporate} account for user registration status. Let F(s)(p) designate the set of users with high graphics and multimedia preferences for whom a registration status is assigned, i.e., those who have already visited the Web site. For the sake of simplicity and without loss of generality, let us consider that in general the site visitors set referred above as functional F, is dependent upon browser settings (v), data access device type (e), personal preferences and access rights: F=F((v), (e), …). In this case, the formula $||F=F((v), (e), …)||$ indicates a formal procedure that evaluates parameterized functional, the expression $||F=F((v), (e), …)(s)||$ evaluates users with given customary settings (s), and the formal procedure $||F=F((v), (e), …)(s)(p)||$ evaluates users with given customary settings (s) and registration status (p). The introduced functional F can be considered an illustration of computational formalism for parameterized procedure of profile evaluation of certain visitor categories.

Let us demonstrate that two-level conceptualization scheme is sufficient for the model adequacy. Let us introduce the following denotations:

$||r|| = \{r_{c.s.}, r_{r.s.}\}$ – specific costs;

$||z|| = \{z_{c.s.}, z_{r.s.}\}$ – segmentation degree (i.e., possibility of splitting users into stable and independent groups);

$||q_i|| = q_i$ – overheads;

$||l_i|| = l_i$ – duration of the request processing stage (download, dynamic form or report creation, etc.);

$||n_i|| = n_i$ – number of request processing stages.

Evaluated values are generalized, i.e., there is no uniqueness of value choice for specific costs and segmentation degree. Generalization level decrease is achieved by considering an assignment point s:

$$||z||(||s||) = \begin{cases} ||z||(higraph) = z_{higraph}, \\ ||z||(mmedia) = z_{mmedia}; \end{cases}$$

$$||r||(||s||) = \begin{cases} ||r||(higraph) = r_{higraph}, \\ ||r||(mmedia) = r_{mmedia}. \end{cases}$$

Moreover, further generalization level decrease by introducing the second assignment p does not succeed:

$||z||(||s||)(||p||) = ||z||(||s||);$

$||r||(||s||)(||p||) = ||r||(||s||).$

The result obtained can be explained by the fact that the evaluation procedure involves visitor position in the structure.

However, it is obvious that overheads $q_i$ are dependent both on customary settings functions and on registration status, i.e. we should let $||q_i|| = \{q_{i\ higraph}, q_{i\ mmedia}\}$. The equality $||q_i|| = q_i$ implies that $q_{i\ higraph} = q_{i\ mmedia} = q_i$.

## 5. The Integrated ERP Implementation

### 5.1 Customizing the Implementation Scheme

During design process, ERP specification is transformed from problem domain concepts to data model entities, then, further, to DBMS scheme (with *PL\SQL* as DO manipulation language), and, finally, to target ERP description with required architecture and interface (see fig. 1). As a result of problem domain analysis, computational DM and generalized scheme of ERP development [16] have been customized to satisfy the required problem domain conditions.

Web applications implementation is finalizing ERP implementation stage. *Oracle Internet Developer Suite* serves a gateway between data warehouse and corporate Internet site. When content-critical warehouse updates occur, content is automatically updated accordingly. Periodical and manual data updates and retrievals are also supported options.

### 5.2 Problem-Oriented Interface and Event-Driven Architecture

According to the detailed ERP design sequence, a comprehensive heterogeneous repository processing scheme is introduced that allows users to interact with distributed database in a certain state depending on dynamically activated (i.e., assigned) scripts. Thus, scripts (in a form of data access profiles and stored object-oriented program language procedures) are initiated depending on user-triggered events. Scripts provide transparent and intellectual client/server front-end user-to-database connection. Dynamically adjustable database access profiles provide high fault tolerance and data security both for ordinary and privileged system users in heterogeneous environment. The profiles are implemented using CORBA technology as an intellectual media between end-user and heterogeneous data warehouses. Depending on semantics-oriented user profile structure, certain database connection and access level profiles are dynamically assigned. The profiles are valid only until the end of data exchange session. According to the hierarchy, users access data under one of the basic scenario profiles. Access is granted not only to data, but also to metadata (i.e., data object dimensions, integrity constraints, access rights, browser parameters, user preferences, etc.).

Administrative users have extended access to metadata. Thus, under the model introduced, data and metadata objects are manipulated uniformly. This makes system interface a problem-oriented, straightforward and uniform one and significantly increases system performance. Internet-published warehouse data processing scheme under conditions of event-driven architecture is presented in fig.2. Client-side web page object states can change depending on event script execution. Though the warehouse data remains unchanged, user can request data update, or produce a query. Moreover, front-end interface itself is also client profile dependent. Options include personal preferences (color schemes, screen resolution etc.), data access device and Web browser settings.

The essential benefit of event-orientation for the ERP global network extension is that corporate users get additional access to Intranet resources of corporate Internet site. Registered (and/or extranet) users access some extra data compared to ordinary ones.

Warehouse data access is also dependent on user profile. At the upper level of data access hierarchy, clients can be divided into administrators, managers and ordinary users (see fig.3). Judging by the profile, data and metadata object states (i.e., system interface) are changed. For instance, a web designer rights assignment provides full access to interface elements database, while content manager can get full access to another warehouse instantiation by means of a different interface instance.

Scenario-based end-user interface results in higher degree of interactivity, user-friendliness and security. User profiles (i.e. assignments) can be stored in metadata base of visitors, and, depending on their properties, data access and representation level could be customized. Sets of Web pages accessible and their content are dependent on user profile. Client-side profile also accounts for preferable information layout depending both on personal preferences and on the device type for which the data is customized. User profile and preferences are also vital for client analysis, and resulting statistic reports serve a foundation for performance and interface optimization.

### 5.3 Implementation Description

The introduced methodology has been practically approved during HR ERP system component improvement at *ITERA* International Group of

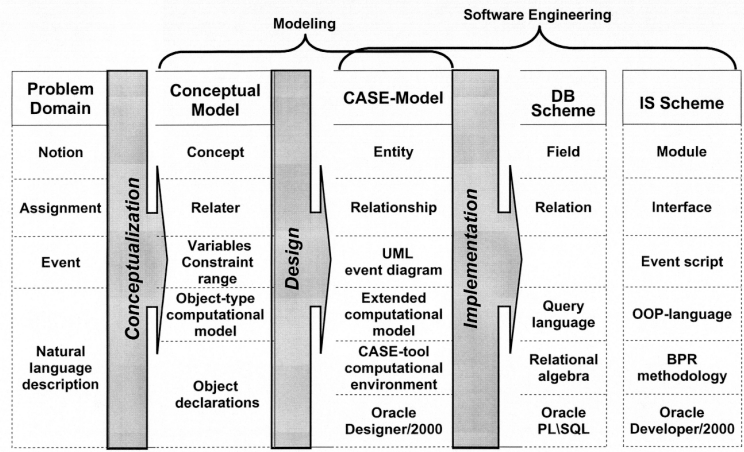

**Figure 1 Generalized ERP implementation chart**

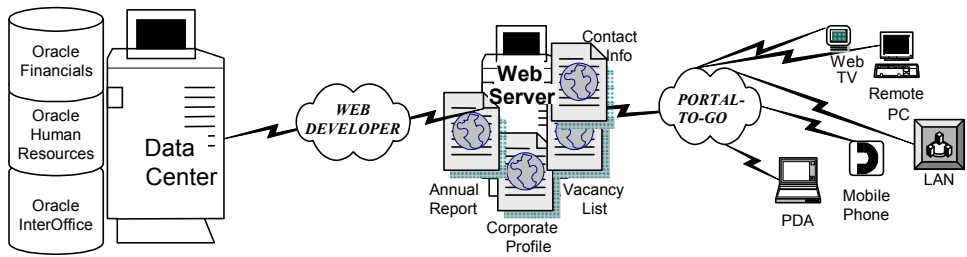

**Figure 2 Event-driven Web-wired ERP data publishing and access scheme**

The Integrated Approach to ERP: Embracing the Web

Companies. All of the HR component modules are cemented together by a unified interface and integrated into ERP environment of *Oracle Applications* family of financial, commodity and document management systems.

From the system architecture viewpoint, the integrated HR component provides certain level of data input, correction, analysis and output depending on front-end position (i.e., assignment) in user hierarchy. Interactive interface is represented by problem-oriented form designer, report generator, on-line documentation and administration tools. The enhanced ERP system database supports the integrated storage for data (i.e., information for on-line users) and metadata (i.e., data object dimensions, integrity constraints and business process parameters). During the enhanced ERP design process, problem domain DM specification (in a semantic network representation) is transformed into use-case UML diagrams, then, by means of *Oracle Developer/2000* and integrated CASE-tool, - into ER-scheme and, finally, into the attributes of target databases. *Oracle Internet Developer Suite* has been used to transform the ERP into a Web-wired application.

On the basis of the information model developed, an architecture and interface solution for integrated HR management software has been designed. To prove adequacy of the model developed and component integration algorithm suggested, software prototype has been designed. To provide required levels of industrial scalability and fault tolerance, judging by the results of exhaustive CASE-and-RAD software analysis, *Oracle Developer/2000* toolkit has been chosen as a solution supporting UML and BPR methodologies. According to specification requirements developed by the author, implementation had been significantly improved by extracting essential information from integrated ERP and publishing it in the corporate Internet site.

HR information subsystem provides a number of significant data items for corporation profile Web page including total establishment, number of countries and companies that represent the corporation. Since the integrated solution implemented includes vacancy module as a part of its HR subsystem, dynamically updated vacancy data Internet page can be easily produced. Similarly, financial components could provide data for a number of periodical or privileged user-triggered financial reports. Data examples include revenues, profits, production dynamics, stock values, etc. Same as with HR subsystem, dynamics tracking is important. Corporation development plans based on deferred charges could be published. Production manufacturing module can provide productivity and capacity data for executive summaries and company profile Web and Intranet pages. Address book from document control subsystem serves for contact information and provides automatic feedback routing through corporate organizational structure. For further enhancement of corporate Web site performance and interface, data published in the Web page is dynamically updated by an event-driven software agent.

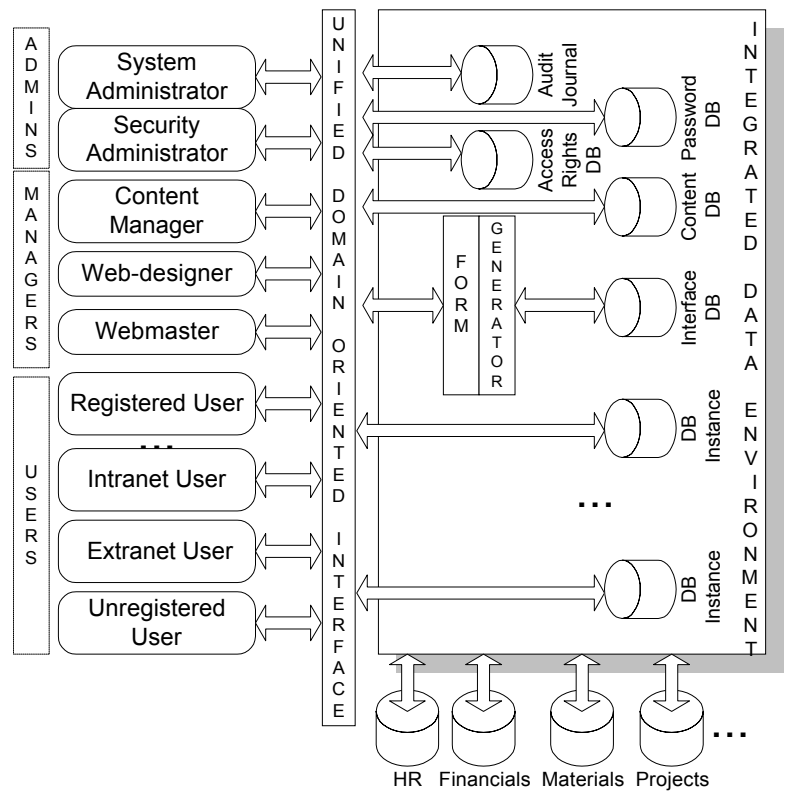

**Figure 3 Front-end interface and data access profiling**

The integrated ERP data warehouse (fig.3) is stored in the data center. ITERA Group warehouse is based on *Oracle Financials, Human Resources* and *InterOffice*. The full-scale implementation is based on hardware platform of an IBM RS/6000 two-server high availability cluster running under IBM AIX operating system. The system has passed a three-year test in a large corporation.

## 6. Results and Conclusion

A computational data model has been introduced that provides integrated manipulation of data and metadata objects, especially in rapidly changing heterogeneous problem domains. The model is an alloy of methods of finite sequences, categories and semantic networks.

On the basis of the formal model, an original and comprehensive iterative scheme for Web-integrated ERP design and implementation has been proposed. The scheme includes an algorithm for new component integration into existing ERP environment. The integration algorithm provides adequacy, consistency and data integrity; algorithm details are presented in [16].

According to the approach suggested, a comprehensive Web-enabled ERP interface has been designed. The ERP interface is based on an open and extendable architecture. As a first step towards implementing the enterprise resource management solution, a fast event-driven software prototype has been developed on the basis of the designed UML-based interface and architecture scheme. Using the prototype testing results, a full-scale object-oriented ERP application has been designed. The full-scale enterprise-level software has been customized for corporate resource management and implemented at an enterprise with more than 1000 employees.

Web-integrated ERP solution have proved significant decrease in time and costs of implementation. Other major benefits include growth of portability, expandability, scalability and ergonomics levels in comparison with existing commercial software of the kind. Iterative multi-level software design scheme is based on formal model unifying object-oriented methods of data (data objects) and knowledge (metadata objects) management. Industrial implementation of the Internet-embracing ERP HR component has been carried out using integrated CASE-and-RAD and Web publishing toolkits. Experience of implementation support has proved importance, urgency, originality and efficiency of the approach suggested. Theoretical and practical statements outlined in the paper have been approved by successful implementation of the Web-wired enterprise-level ERP software at *ITERA* International Group of Companies. The author is going to continue research in order to turn the enhanced ERP into an integrated corporate Intranet and e-commerce solution.